\documentclass[aps,12pt,eqsecnum,showpacs,nofootinbib]{revtex4}
\usepackage{setstack}
\usepackage{amsfonts}
\usepackage{amssymb}
\usepackage{color}

\def\bc{\begin{center}}
\def\nno{\nonumber}
\def\ec{\end{center}}
\def\be{\begin{eqnarray}}
\def\ee{\end{eqnarray}}

\definecolor{dyellow}{rgb}{1.,0.8,.0}
\definecolor{myblue}{rgb}{.1,.1,.7}
\definecolor{dcyan}{rgb}{.0,.6,.6}
\definecolor{dmagenta}{rgb}{0.6,0.0,0.6}
\definecolor{brown}{rgb}{0.6,0.2,0.}
\definecolor{darkblue}{rgb}{.0,.0,0.5}
\definecolor{darkred}{rgb}{0.75,0.0,0.0}
\definecolor{orange}{rgb}{1.,.6,.0}
\definecolor{dorange}{rgb}{0.8,.4,.0}
\definecolor{darkgreen}{rgb}{0.0,0.6,0.0}
\definecolor{purple}{rgb}{.4,.0,.4}
\def\La{\Lambda}
\def\Si{\Sigma}

\def\dl{\delta}

\def\si{\sigma}

\def\d#1#2{\frac{\displaystyle #1}{\displaystyle #2}}

\newcommand\btd{\raise 2pt
\hbox{$\hat\bigtriangledown$}\hskip 1.5pt}
\newcommand\bt{\raise 2pt
\hbox{$\bigtriangledown$}\hskip 1.5pt}

\newcommand{\BdS}{${B}d{S}$}
\newcommand{\dS}{$d{S}$}
\newcommand{\AdS}{${A}d{S}$}
\newcommand{\Mink}{$Mink$}

\newcommand{\PoI}{${P}o{I}$}
\newcommand{\CP}{cosmological principle}%$CP$}

\newcommand{\SR}{special relativity}%${\cal SR}_c$}

\newcommand{\F}{${FLT}$}
\newcommand{\GR}{general relativity}%${\cal GR}$}
\newcommand{\omits}[1]{}

\def\PRD{{\it Phys. Rev.}~{\bf D}}

\def\PLA{{\it Phys. Lett.}~{\bf A}}

\def\CTP{{\it Commun. Theor. Phys. }}

\begin{document}
\title{On Principle of Inertia in Closed Universe}

\author{Han-Ying Guo$^{1,2}$\footnote{hyguo@itp.ac.cn}}

\affiliation{
    CCAST (World Laboratory), P.O. Box 8730, Beijing 100080, China,}
\affiliation{
    Institute of Theoretical Physics, Chinese Academy of Sciences,
    %P.O. Box 2735,
    Beijing 100080, China}

\begin{abstract}
If our universe is asymptotic to a  de Sitter space, it should be
closed
 with  curvature in $O(\Lambda)$ in view of  \dS\, \SR. Conversely,
 its evolution  can fix on  Beltrami systems of inertia in the
\dS-space without {Einstein's} `argument in a circle'. Gravity
should be local \dS-invariant based on localization of the principle
of inertia.
\end{abstract}

\pacs{04.20.Cv %Fundamental problems and general formalism
%04.90.+e, %Other topics in general relativity and gravitation (restricted to new topics in section 04 \footnote{Section 04 is General relaivity and Gravitation)
%04.50+h, %Gravity in more than four dimensions, Kaluza-Klein theory, unified field theories; alternative theories of gravity
98.80.Jk, %Mathematical and relativity aspect of cosmology
03.30.+p, %special relativity
02.40.Dr, %Euclidean and projective geometry
}

\maketitle%
\tableofcontents

%%%%%%%%%%%%%%%%%%%%%%%%%%%%%%%%%%%%%%%%%%%%%%%%%%%%%%%%%%%%%%%%%%%%%%%%
%%%%%%%%%%%%%%%%%%%%%%%%       Section 1        %%%%%%%%%%%%%%%%%%%%%%%%
%%%%%%%%%%%%%%%%%%%%%%%%%%%%%%%%%%%%%%%%%%%%%%%%%%%%%%%%%%%%%%%%%%%%%%%%
%%%%%%%%%%%%%%%%%%%%%%%     Introduction        %%%%%%%%%%%%%%%%%%%%%%%%
\section{Introduction}

{In classical physics,} it is well known that for  both Newton
theory and Einstein's \SR\, the principle of inertia (\PoI) {with
Galilean symmetry and Poincar\'e symmetry, respectively,} plays an
extremely important role as the benchmark of physics {for defining
physical quantities and introducing physical laws}. But, in
Einstein's point of view, there is an `argument in a circle' {for
the \PoI\ as the benchmark}.

Some eighty five years ago,  Einstein claimed:

{\it `The weakness of the \PoI\, lies in this, that it involves an
argument in a circle: a mass moves without acceleration if it is
sufficiently far from other bodies; we know that it is sufficiently
far from other bodies only by the fact that it moves without
acceleration. Are there at all any inertial systems for very
extended portions of the space-time continuum, or, indeed, for the
whole universe? We may look upon the principle of inertia as
established, to a high degree of approximation, for the space of our
planetary system, provided that we neglect the perturbations due to
the sun and planets. Stated more exactly, there are finite regions,
where, with respect to a suitably chosen space of reference,
material particles move freely without acceleration, and in which
the laws of the special theory of relativity, $\cdots$, hold with
remarkable accuracy. Such regions we shall call ``Galilean
regions."' }\cite{Einstein23}

In fact,  to avoid this `weakness' is one of the main motivations
for Einstein from \SR\ to \GR\, based on his equivalence principle
and general principle of relativity as an extension of the special
principle of relativity.

In \GR, however, what is realized for the general principle of
relativity is the principle of general covariance. Although it is
always possible to analyze physics in terms of arbitrary
(differentiable) coordinate systems  at classical level, `the
principle of covariance has no forcible content.'\cite{MTW}
  For the equivalence principle, it
  requires that
  physical quantities and laws are in `their familiar special-relativistic
  forms'  in  local Lorentz
  frames \cite{MTW}. The symmetry for physical
  quantities and laws, however, is local $GL(4,R)$ or its subgroup
  $SO(1,3)$ without local translation in general. Thus, in `Galilean
  regions',
   Poincar\'e symmetry of  \PoI\, as
  the benchmark in special relativity is partially lost. These seem away from
  Einstein's original intention {and lead to the benchmark of physics with gravity
  is not completely in consistency with that in \SR\ without gravity}.

Recent observations show that our universe is accelerated
expanding\cite{Riess, WMAP}. It is certainly not asymptotic to a
Minkowski (\Mink)-space, rather  quite possibly asymptotic to a de
Sitter (\dS)-space with a tiny  cosmological constant $\La$. These
present great challenges to the foundation of physics on the cosmic
scale (see, e.g., \cite{Witten}). In fact, it is the core of
challenges: {What are the benchmarks of physics  on the cosmic
scale? Are they consistent?}

In view of the \dS-invariant \SR\,\cite{Lu, LZG, BdS, BdS2, Lu05,
TdS, IWR, NH, C3}, however, there is a \PoI\, {of \dS-invariance}
 on \dS-space with  Beltrami systems of inertia (denoted \BdS-space). Here we
 show that
 if the universe is asymptotic to  a \dS-space, it should be closed
 with a tiny curvature in the order of $\Lambda$,  $O(\Lambda)$. Conversely,
 the evolution of
the universe can fix on the Beltrami systems. Thus, the universe
acts as the origin of the \PoI\, {of \dS-invariance} without
Einstein's `argument in a circle' so that the benchmark of physics
on the cosmic scale should still be the \PoI\, {of \dS-invariance}.
Then, we explain that the benchmark of physics with gravity should
be the localization of the \PoI\, of the \dS\, \SR. Thus, the \PoI\,
{of \dS-invariance} and its localization should play the role of the
{consistent} benchmark{s} of physics { on the cosmic scale} in the
universe.

Actually, based on the principle of relativity  \cite{Lu, LZG} and
the postulate on invariant universal constants, the speed of light
$c$ and the curvature radius $R$ \cite{BdS, BdS2}, the \dS\,
special relativity can be set up on the \BdS-space. While
Einstein's \SR\ is the limiting case of $R \to \infty$.

In  the  \dS\, \SR,  Beltrami coordinate systems \cite{beltrami}
with
 Beltrami time simultaneity are very similar to  \Mink-systems in
Einstein's \SR. Namely, in the \BdS-space  geodesics are all {\it
straight world lines} so that there is a \PoI\, with a {\it law of
inertia} for  free particles and light signals. All these issues are
transformed symmetrically under the fractional linear
transformations with a common denominator (\F s) of  \dS-group
$SO(1,4)$ in the Beltrami atlas chart  by chart. It is significant
that the Beltrami systems and their Robertson-Walker-like
\dS-counterpart with respect to  proper-time simultaneity provide an
important model. In this model, the \dS-group
 as a maximum symmetry
ensures that there are both the \PoI\ and the cosmological principle
on  \dS-space as two sides of a coin. On one side, there is the
\BdS-space
 with the \PoI, while on the other there is a Robertson-Walker-like
 \dS-space with the \CP\, having an accelerated
expanding closed 3-d cosmos $S^3$ of curvature in the order of
$O(R^{-2})$. Since the both can be transformed each other explicitly
by changing the simultaneity just like flip a coin, the
Robertson-Walker-like \dS-space   displays as an origin of the \PoI,
while the \PoI\, provides a benchmark of physics on the \dS-space.

If the universe is asymptotic to  a \dS-space with
$R\simeq(3/\Lambda)^{1/2}$. In  view of the \dS\ \SR, the universe
should be asymptotic to the Robertson-Walker-like \dS-space in the
model so that it should be closed and the deviation from flatness is
in the order of $\Lambda, O(\Lambda)$. This is an important
prediction
 more or less  consistent with  recent data from WMAP
\cite{WMAP} {and can be further checked}.

Conversely, the asymptotic behavior of the universe should naturally
pick up  a kind of the Robertson-Walker-like \dS-systems with such a
`cosmic' time $\tau$ that its axis coincides with the revolution
time arrow of the real cosmic time $\tau_u$ in the universe. Since
the `cosmic' time $\tau$ in the Robertson-Walker-like \dS-space is
explicitly related to the Beltrami time $x^0$, the universe should
also fix on a kind of
 Beltrami systems with  $x^0$ transformed from
the `cosmic' time $\tau$. Therefore, via its evolution time arrow of
$\tau_u$ picking up a `cosmic' time $\tau$ on the
Robertson-Walker-like \dS-space, the universe should just act as an
origin of such  kind of Beltrami systems in which the \PoI\ holds.
Thus, there do exist the inertial systems in the universe and there
is no  Einstein's `argument in a circle' for the \PoI.

In \GR, there is no special relativity in \dS-space. In the \dS\,
\SR, there is no gravity in \dS-space.
 How to describe gravity?

 In the
 light of Einstein's `Galilean regions' \cite{Einstein23}, where his
 \SR\ with full Poincar\'e symmetry should hold locally,  the \PoI\,
 should be localized. Therefore, in view of the  \dS\, \SR,  on spacetimes with
 gravity there should be local
\dS-frame anywhere and anytime so that the \PoI\, of the  \dS\,
\SR\, should hold
  locally. If so, the localized \PoI\ of the  \dS\, \SR\, should be the benchmark
  of physics with gravity. This is in consistency
  with the role played by the \PoI\, of the  \dS\, \SR.
  We may further require that gravity  have  a
gauge-like dynamics characterized by a dimensionless constant
$g\simeq (\Lambda G\hbar /c^{3})^{1/2}\sim 10^{-61}$ from the
cosmological constant $\Lambda$ and the Planck length. A simple
model has implied this should be the case \cite{dSG, hyg76, T77}.

This letter  is arranged as follows. In  sections 2, we argue why
there is a \PoI\, on \dS-space and very briefly introduce the \dS\,
\SR. In section 3, we introduce the relation between the \PoI\, and
the \CP\, on \dS-space as well as the cosmological meaning of \dS\,
\SR. In section 4, we explain why the universe can fix on the
Beltrami systems of inertia without Einstein's `argument in a
circle'. In section 4, we very briefly discuss that gravity should
be based on  localization of the  \dS\, \SR\, with \PoI\,  and
introduce the simple model for  \dS-gravity. Finally, we end with a
few remarks.

%%%%%%%%%%%%%%%%%%%%%%%%%%%%%%%%%%%%%%%%%%%%%%%%%%%%%%%%%%%%%%%%%%%%%%%%
%%%%%%%%%%%%%%%%%%%%%%%%       Section 2       %%%%%%%%%%%%%%%%%%%%%%%%
%%%%%%%%%%%%%%%%%%%%%%%%%%%%%%%%%%%%%%%%%%%%%%%%%%%%%%%%%%%%%%%%%%%%%%%%
%%%%%%%%%%%%%%%%%%%%%%%  On dS invariant SR    %%%%%%%%%%%%%%%%%%%%%%%%

\section{On de Sitter special relativity}

Is there  \SR\, with a \PoI\, on \dS-space?

Yes! Absolutely. This can be enlightened from two deferent but
related angles \cite{Lu, LZG, BdS, BdS2, Lu05, TdS, IWR}.

Firstly, as is well known, weakening the Euclid fifth axiom leads to
Riemann and Lobachevsky geometries on an almost equal footing with
Euclid geometry. There is a physical analog  via  an inverse Wick
rotation of  4-d Euclid space, Riemann sphere and Lobachevsky
hyperboloid $E^4/S^4/L^4$, respectively. Namely, there should be two
other  kinds of the \dS/\AdS-invariant \SR\, on an almost equal
footing with Einstein's special relativity \cite{IWR}. In fact,
there is a one-to-one correspondence between three kinds of
geometries and their
physical counterparts. We list the correspondence as  follows:%
 \bc
 \begin{tabular}{lcl}
{ \quad Geometry}  &  ~~& { Spacetime Physics}\\
${E}^4/{S}^4/{L}^4$  &  &   ${M}^{1,3}$/\dS$^{1,3}$/\AdS$^{1,3}$\\
$ISO(4)/SO(5)/SO(1,4)$& &  $ISO(1,3)/SO(1,4)/SO(2,3)$\\
\rm{Descartes,~Beltrami atlas} & &
\rm{Minkowski,~ Beltrami atlas}\\
Points &  & Events\\
Straight line&  & Straight world line \\
Principle of Invariance & &   Principle of Relativity\\
Klein's Erlangen Programm  &  &   Theory of Special Relativity\\
\end{tabular}\ec

Secondly, owing to Umov, Weyl and Fock \cite{Fock}, it can be proved
that the most general form of the transformations among inertial
coordinate systems%
\be\label{FL}%
{x'}^i=f^i( x^i), \quad x^0=ct, \quad i=0,\cdots, 3,%
\ee
which transform a uniform straight line motion, i.e. the inertial
motion, in $F(x)$%
\be\label{imt}%
x^a=v^a (t-t_0)+x^a_0,~~~ ~v^a=\frac{dx^a}{dt}={\rm const.}\quad a=1,2,3,%
\ee%
 to a motion of the same nature in $F'(x')$, are of  \F-type.

As in Einstein's \SR, the principle of relativity implicates that
there is a metric in inertial  systems on 4-d spacetime with
signature $\pm 2$ and it is invariant under a transformation  group
with ten parameters including spacetime `translations', boosts and
space rotations. Thus, these 4-d spaces are maximally symmetric,
i.e.
 \Mink/\dS/\AdS\, of zero,
positive  or negative constant curvature, invariant under group
 $ISO(1,3)/SO(1, 4)/SO(2, 3)$, respectively. As for invariant universal
 constants, in addition to the speed of light $c$ there is another invariant
constant $R$, the  radius of \dS/\AdS-spaces. Therefore, the
\dS/\AdS\, \SR\ can be set up based on the principle of relativity
and the postulate on invariant universal constants \cite{BdS, BdS2}.

The \dS-space  as a $4$-d hyperboloid ${\cal H}_R$ can be embedded
in a $1+4$-d \Mink-space, ${\cal H}_R \subset M^{1,4}$:
 \be\label{5sphr}%
 {\cal H}_R:  &&\eta^{}_{AB} \xi^A \xi^B= -R^2,
\\ %
\label{ds2}%
&&ds^2=\eta^{}_{AB} d\xi^A d\xi^B, %~
\ee %
where $\eta_{AB}= {\rm diag}(1, -1, -1, -1, -1)$, $A, B=0, \ldots,
4$.

On the hyperboloid, a kind of uniform great `circular' motions of a
particle with mass
$m_R$ can be defined by a conserved 5-d angular momentum:%
\be\label{nmonH}%
\frac{d{\cal L}^{AB}}{ds}=0,\qquad {\cal
L}^{AB}:=m_R(\xi^A\frac{d\xi^B}{ds}-\xi^B\frac{d\xi^A}{ds}).%
\ee%
For the particle, there is an Einstein-like formula:%
\be\label{massH}%
-\frac{1}{2R^2}{\cal L}^{AB}{\cal L}_{AB}=m^2_R, \qquad {\cal
L}_{AB}:=\eta_{AC}\eta_{BD}{\cal L}^{CD}.
\ee%
Obviously, the eqns (\ref{5sphr}), (\ref{ds2}), (\ref{nmonH}) and
(\ref{massH}) are invariant under linear transformations of
\dS-group $SO(1,4)$. For  a massless particle or a light signal with
$m_R=0$, similar motion can be defined as long as the proper time
$s$ is replaced by an affine parameter $\lambda$.

  Via a `gnomonic'
projection without antipodal identification, ${\cal H}_R$ becomes
the \BdS-space with a Beltrami atlas \cite{BdS, BdS2} chart by
chart. In the charts $U_{\pm 4}$, for instance,
\be \label{u4}%
&&x^i|_{U_{\pm 4}}
=R {\xi^i}/{\xi^4},\qquad i=0,\cdots, 3;  \\
&&\xi^4|_{U_{\pm 4}}=({\xi ^0}^2-\sum _{a=1}^{3}{\xi ^a}^2+ R^2
)^{1/2} \gtrless 0,
\ee%
 there are condition from (\ref{5sphr}) and
\BdS-metric from (\ref{ds2})
\begin{eqnarray}\label{domain}
\sigma(x)=\sigma(x,x):=1-R^{-2} \eta_{ij}x^i x^j>0,\qquad\\\label{bhl}%
 ds^2=[\eta_{ij}\sigma(x)^{-1}+ R^{-2} \eta_{ik}\eta_{jl}x^k x^l
\sigma(x)^{-2}]dx^i dx^j,
\end{eqnarray}
where $\eta_{ij}%
={\rm diag} (1, -1,-1,-1)$. Under
  \F s of $SO(1,4)$ sending an event $A(a^i)$ to the origin %
\begin{equation}\label{G}
\begin{array}{l}
x^i\rightarrow \tilde{x}^i=\pm\sigma(a)^{1/2}\sigma(a,x)^{-1}(x^j-a^j)D_j^i,\\
\quad D_j^i=L_j^i+{ R^{-2}}%
\eta_{jk}a^k a^l (\sigma(a)+\sigma(a)^{1/2})^{-1}L_l^i,\\
\quad L:=(L_j^i)_{i,j=0,\cdots,3}\in SO(1,3),
\end{array}\end{equation}
  (\ref{domain}) and (\ref{bhl}) are invariant.  Thus,
inertial systems and inertial motions transform among themselves,
respectively.

For a pair of events ($A(a^i)$, $X(x^i)$),
\be\label{lcone0} %
{\Delta}_R^2(a, x) = R^2
[\sigma^2(a,x)-\sigma(a)\sigma(x)] \gtreqqless 0% 
\ee %
is invariant under (\ref{G}). Thus, the pair is time-like, null, or
space-like, respectively.

The Beltarmi light-cone at an event $A$ with running points $X$ is
\be \label{nullcone} %
{\cal F}_{R}:= R
\{\sigma(a,x) - [\sigma(a)\sigma(x)]^{1/2}\}=0.%
 \ee%
At the origin %
$a^i=0$, it is just Minkowskian $\eta_{ij}x^i x^j = 0$.

Under the `gnomonic' projection, the uniform great `circular'
motions are projected as a kind of inertial motions along geodesics.
In fact, the geodesics are Lobachevsky-like straight world lines and
vise versa. A time-like geodesic, along which a particle with mass
$m_{R}$ moves, is equivalent to
\begin{equation}\label{pi}
\frac{dp^i}{ds}=0, \quad p^i:=m_{R
}\sigma(x)^{-1}\frac{dx^i}{ds}=C^i={\rm const.}
\end{equation}
 Under certain initial condition it is just a straight world line
with respect to $w = w(s)$
\begin{equation}
x^i(w)=c^iw+b^i.
\end{equation}%
A light signal moves along a null geodesic with an affine parameter
$\lambda $ can be written as%
\begin{equation}\label{ki}
\frac{dk^i}{d\lambda}=0, \quad k^i:=\sigma
^{-1}(x)\frac{dx^i}{d\lambda }={\rm const.}
\end{equation}
It can also be expressed as a straight line\cite{BdS, BdS2}.

From both (\ref{pi}) and (\ref{ki}), it follows that the coordinate velocity components are constants%
\be\label{im}%
\frac{dx^a}{dt}=v^a={\rm const.}, \quad a=1,2,3.%
\ee%
Thus, the both motions of  free particles and light signals are
indeed of inertia as in (\ref{imt}), chart by chart. Namely, {\it
the law of inertia} holds on the \BdS-space.  This together with the
principle of relativity is just the \PoI\, in the \BdS-space.

For such a free massive particle
 a set of conserved
 quantities $p^i$ in (\ref{pi}) and $L^{ij}$ can be defined as a pseudo
4-momentum vector  and a pseudo 4-angular-momentum, respectively
\begin{equation}\label{angular4}
L^{ij}=x^ip^j-x^jp^i,\qquad \frac{dL^{ij}}{ds}=0.
\end{equation}
 In fact, $p^i$ and
$L^{ij}$ constitute the conserved 5-d angular momentum in
(\ref{nmonH}). And the Einstein-like formula (\ref{massH}) becomes
a {\it generalized
Einstein's formula}%
\begin{eqnarray}\label{eml}%
 E^2=m_{R}^2c^4+{p}^2c^2 + \frac{c^2} {R^2} { j}^2 - \frac{c^4}{R^2}{
k}^2,%
\end{eqnarray}
with energy $E=p^0$, momentum $p^a$, $p_a=\delta_{a b}p^b$, `boost'
$k^a$, $k_a=\delta_{ab}k^b$ and 3-angular momentum $j^a$,
$j_a=\delta_{ab} j^b$. For a massless particle or a light signal
with $m_R=0$, similar issues hold so long as the proper time  is
replaced by an affine parameter.

 If we introduce the Newton-Hooke constant
$\nu$
\cite{NH} and take $R$ as $R\simeq (3/\La)^{1/2}$,%
\be\label{NHc}%
\nu:=\d {c} {R}\simeq c (3/\La)^{-1/2},\quad \nu^2 \sim 10^{-35}s^{-2}. %
\ee%
It is so tiny that all experiments at ordinary scales cannot
distinguish  the \dS\, \SR\  from Einstein's one.

 In order to make
sense  of inertial motions and these observables for an inertial
observer ${\cal O}_I$ rested at the spacial origin of the Beltrami
system, the simultaneity should be defined. Similar to Einstein's
\SR,
 two events $A$ and $B$ are simultaneous if and only if
their Beltrami temporal coordinate values $x^0(A)$ and $x^0(B)$ are
equal:
\be %
a^0:=x^0(A) =x^0(B)=:b^0. %
\ee
It is called the {\it Beltrami simultaneity} and defines a 1+3
decomposition of the \BdS-metric (\ref{bhl})% on \BdS-space
\be%
 ds^2 =  N^2 (dx^0)^2 - h_{ab} (dx^a+N^a dx^0)
(dx^b+N^b dx^0) %
\ee %%
with  lapse function, shift vector, and induced 3-geometry on
 $\Si_c$ in the  chart, respectively,
\begin{eqnarray}
& & N=\{\si_{\Si_c}(x)[1-(x^0 /R)^2]\}^{-1/2}, \nonumber \\%
& & N^a=x^0 x^a[ R^2-(x^0)^2]^{-1},
 \\
& & h_{ab}=\dl_{ab} \si_{\Si_c}^{-1}(x)-{ [R\si_{\Si_c}(x)]^{-2}
\dl_{ac} \dl_{bd}}x^c x^d ,\nonumber
\end{eqnarray}
where $\si_{\Si_c}(x)=1-(x^0{/R})^2 + {\dl_{ab}x^a x^b /R^2}$.
 In particular, at $x^0=0$, $\si_{\Si_c}(x)=1+{ \dl_{ab} x^a
x^b/R^2}$,  3-hypersurface $\Si_c$ is isomorphic to an $S^3$ in all
Beltrami coordinate charts.

%%%%%%%%%%%%%%%%%%%%%%%%%%%%%%%%%%%%%%%%%%%%%%%%%%%%%%%%%%%%%%%%%%%%%%%%
%%%%%%%%%%%%%%%%%%%%%%%%       Section 3      %%%%%%%%%%%%%%%%%%%%%%%%
%%%%%%%%%%%%%%%%%%%%%%%%%%%%%%%%%%%%%%%%%%%%%%%%%%%%%%%%%%%%%%%%%%%%%%%%
%% Proper-time simultaneity and Robertson-Walker-like \dS-space %%%%
\section{Principle of inertia and cosmological principle as two
sides of a coin}

On the \dS-space, there is an important relation between the \PoI\,
and the cosmological principle. It is just like two sides of a coin.

In fact, for an observer rest at  spacial origin $x^a=0$ of Beltrami
system, there is another
 simultaneity: the  {\it proper-time simultaneity}
  with respect to an ideal
clock's proper-time $\tau$. It is easy to see that the proper-time
$\tau$ is explicitly related to the Beltrami  time  $x^0$:
\begin{eqnarray}\label{ptime}
\tau:=\tau_R=R \sinh^{-1} (R^{-1}\sigma^{-\frac{1}{2}}(x)x^0).
\end{eqnarray}%
Thus,  the  {\it proper-time simultaneity}  can be defined as: all
events $X(x^i)$ are  simultaneous with respect to %the proper-time of
the observer if and only if their proper time are equal. Namely,
\begin{equation}\label{smlt}
x^0\sigma^{-1/2}(x,x)=:\xi^0=R \sinh(R^{-1}\tau)=\rm constant.%  
\end{equation}
In fact, these events are comoving with the observer, who now
becomes a comoving one ${\cal O}_C$ with respect to all these
events.
The line-element %
on a simultaneous  hypersurface ${\Sigma_\tau}$ now is
\begin{equation}\label{dl}
dl^2=-ds^2_{\Sigma_\tau},%
\end{equation}
where
\be \begin{array}{l}\label{spacelike}
ds^2_{\Sigma_\tau} = R_{\Si_\tau}^2%
dl_{{\Sigma_\tau} 0}^2, \\
R_{\Sigma_\tau}^2%
:=\sigma^{-1}(x,x)\sigma_{\Sigma_\tau}(x,x)%
=1+ (\xi^0/R)^2,\\%
\sigma_{\Sigma_\tau}(x,x):=1+R^{-2}\delta_{ab}x^a x^b
>0, \\ %
dl_{{\Sigma_\tau} 0}^2:={ \{\delta_{ab}\sigma_{\Sigma_\tau}^{-1}(x)
-[R\sigma_{\Sigma_\tau}(x)]^{-2}\delta_{ac}\delta_{bd}x^c x^d\}}
 dx^a dx^b.
\end{array}\ee

It is clear that this simultaneity is directly related to the
cosmological principle on the \dS-space. In fact,
 if the proper time $\tau$ is taken as
a temporal coordinate for the  observer ${\cal O}_C$, the
\BdS-metric (\ref{bhl}) becomes as a Robertson-Walker-like
\dS-metric with $\tau$ being a `cosmic' time and an accelerated
expanding 3-d cosmos isomorphic to $S^3$:
\begin{equation}\label{dsRW}
ds^2=d\tau^2-dl^2=d\tau^2-\cosh^2( R^{-1}\tau) dl_{{\Sigma_\tau} 0}^2.%
\end{equation}

It is important that two kinds of simultaneity relate the
\BdS-metric (\ref{bhl}) with the \PoI\, and the
Robertson-Walker-like \dS-metric (\ref{dsRW}) with the cosmological
principle. They do make sense in two types of measurements: the
Beltrami simultaneity is for those of the inertial observer ${\cal
O}_I$ relevant to the \PoI\,  and the proper time simultaneity for
those  of the comoving observer ${\cal O}_C$ concern `cosmic'
effects of all distant stars and cosmic objects except the
cosmological constant as test stuffs. Thus, on the \dS-space there
is a kind of  inertial-comoving observers ${\cal O}_{I-C}$ who play
two roles with apparatus having two different types of time scales
and relevant rulers. What should be done for them from their
comoving observations to another type of measurements is to switch
off the `cosmic' time $\tau$ with the `cosmic' rule and on the
Beltrami time $x^0=ct$ with the Beltrami rule, respectively, and
vise versa. Namely, if the observers as comoving ones, ${\cal
O}_{C}$, on (\ref{dsRW}) would change their measurements from the
proper-time simultaneity to the Beltrami time one according to the
relation (\ref{ptime}), they become inertial ones ${\cal O}_{I}$,
for whom the \PoI\, makes sense, and vise versa.

Actually,  for the \dS-space this provides
  a very meaningful model  like a coin with two sides. On one side,
  there is the \PoI\, on
  the \BdS-space (\ref{bhl}) together with
   the law of inertia on
inertial systems with respect to a set of inertial observers ${\cal
O}_{I}$. On another side, the Robertson-Walker-like \dS-space
(\ref{dsRW}) displays the \CP\, with respect to a set of comoving
observers ${\cal O}_{C}$. In other words, the `cosmic' background of
the Robertson-Walker-like \dS-space (\ref{dsRW})  supports the
\PoI\, on the \BdS-space  (\ref{bhl}). And conversely, the \PoI\,
provides a benchmark of physics related to  `cosmic' observations.

%%%%%%%%%%%%%%%%%%%%%%%%%%%%%%%%%%%%%%%%%%%%%%%%%%%%%%%%%%%%%%%%%%%%%%%%
%%%%%%%%%%%%%%%%%%%%%%%%       Section 4        %%%%%%%%%%%%%%%%%%%%%%%%
%%%%%%%%%%%%%%%%%%%%%%%%%%%%%%%%%%%%%%%%%%%%%%%%%%%%%%%%%%%%%%%%%%%%%%%%
%%%%%%%%%%%%%%%%%%%%%%%%%%%   determination of PoI  %%%%%%%%%%%%%%%%%%%%%%%%%%%%%
\section{Are there any inertial systems for the whole universe?}

`Are there at all any inertial systems for very extended portions of
the space-time continuum, or, indeed, for the whole universe?
'\cite{Einstein23} For Einstein, the answer seems to be negative
unless for the `Galilean regions'. However, in view of the  \dS\,
\SR, the answer is positive!

Actually, the universe does fix on a kind of inertial systems in the
following manner. Firstly, if the universe is accelerated expanding
and asymptotic to a \dS, its fate should be the
Robertson-Walker-like \dS-space (\ref{dsRW}). This is very natural
in view of the \dS\ \SR.
  Secondly, the time direction and the homogeneous space of the universe tend to the `cosmic' time and the 3-d  cosmos as an
accelerated expanding $S^3$ of the Robertson-Walker-like \dS-space,
respectively. These set up the directions of the `cosmic' time axis
and the spacial axes for the Robertson-Walker-like \dS\,  up to some
spacial rotations in all them  transformed each other by \dS-group.
Thirdly, by means of the important relation
 between the \BdS-metric (\ref{bhl})
and the Robertson-Walker-like \dS-metric (\ref{dsRW}) by changing
the simultaneity, or just simply via the relation (\ref{ptime})
between the Beltrami time $x^0$ and the `cosmic' time $\tau$,
 the directions of the axes of the
Beltrami systems can be given. In fact, the Beltrami time axis is
related to the `cosmic' time axis in the Robertson-Walker-like
\dS-space, while the spacial axes of the Robertson-Walker-like
\dS-metric (\ref{dsRW}) are just the Beltrami spacial ones in the
\BdS-metric (\ref{bhl}). Thus, the evolution of the universe does
fix on the Beltrami inertial systems.

 It is important that
such a way of determining the Beltrami systems of inertia is
completely different from the way of Einstein  \cite{Einstein23}.
Actually, the gravitation in the universe does not explicitly play
any roles here and there is nothing related to Einstein's `argument
in a circle.'

  In the Beltrami systems, there are two  universal constants, $c$ and $R$.
In order to set up the Beltrami systems, it is also needed to
determine their values concretely. However, it is  clear that as
inertial-frames the Beltrami systems do not depend on their
concrete values unless they are related to observations in the
universe. In this case, their values should be given by two
independent experiments or observations. Note that these constants
are supposed to be invariant and universal approximately. So, the
speed of light $c$ may still be taken as that in Einstein's \SR,
which is just a limiting case $R\to \infty$ of the \dS\, \SR.
Thus,  this also fixes on
 the origin of the Beltrami systems since
 the Beltrami light cone (\ref{nullcone}) at the origin is just Minkowskian.
 As for  the value of $R$, it %And the value of $R$
may also be given by $R\simeq(3/\Lambda)^{1/2}$ with the $\Lambda$
being taken in the precise cosmology nowadays. Furthermore, the
re-scaling of the curvature radius $R$ may lead to the conformal
extension and compactification of the \dS-space together with that
of the \Mink-space and the \AdS-space \cite{C3}.

It is also clear and important that although the  temporal axis of
such kind of Beltrami systems  can be fixed on by the evolution  of
the universe in the above manner, the symmetry among all Beltrami
systems is still of the \dS-group so long as the cosmological
effects are not be taken into account. Otherwise, the symmetry
should be reduced to the group $SO(4)$ for the comoving observations
in the universe. This may shed light on the inconsistency between
the principle of relativity and the cosmological environment (see,
e.g. \cite{BBR}).

Further, different  kinds of  \PoI\, together with relevant
inertial-frames in all possible kinematics, such as  Einstein's \SR,
Newton mechanics, Newton-Hooke mechanics \cite{NH} and so on can be
viewed as certain contractions in different limits of $c$ and $R$,
respectively. Therefore, the origin of all these \PoI\, should be
inherited from the \PoI\, in the \BdS-space and in this sense they
can also be set up
by the evolution of the universe. % without Einstein's `argument in a circle'.

In conclusion, the Beltrami systems of inertia and their
contractions does exist in the universe. Their coordinate axes can
be fixed on by the cosmic time's arrow of the universe via the
Robertson-Walker-like \dS-space, to which the universe is
asymptotic. This is independent of  the gravitational effects. In
this sense, for the \PoI\,  in the  \dS\, \SR\, and all other kinds
of \PoI\, as its contractions,  there is no longer Einstein's
`argument in a circle' \cite{Einstein23}.

Of course, in the universe except at its fate as a \dS-space,
there is gravity anywhere and anytime. How to take into account
the gravitational effects and what should be done for the \PoI?
What is the benchmark of physics with gravity?

%%%%%%%%%%%%%%%%%%%%%%%%%%%%%%%%%%%%%%%%%%%%%%%%%%%%%%%%%%%%%%%%%%%%%%%%
%%%%%%%%%%%%%%%%%%%%%%%%       Section 5       %%%%%%%%%%%%%%%%%%%%%%%%
%%%%%%%%%%%%%%%%%%%%%%%%%%%%%%%%%%%%%%%%%%%%%%%%%%%%%%%%%%%%%%%%%%%%%%%%
%%%%%%%%%%%%%%%%%%%%%%%%%%%     Gravity       %%%%%%%%%%%%%%%%%%%%%%%%%%%%%
\section{Gravity and localized principle of inertia}

In view of  the  \dS\, \SR, there is no gravity in the \dS-space.
The `gravitational effects' in the \dS-space with coordinate atlas
other than the Beltrami one should be a kind of non-inertial
effects. Temperature and entropy in the static \dS-system are just
this case in analogy with the Rindler space in view of Einstein's
\SR\, in the \Mink-space\cite{TdS}. Thus, the \dS-space does not
like a black hole.

In order to describe gravity, we would like to recall Einstein's
description on  `Galilean regions' first. In these finite regions,
`the laws of the special theory of relativity, $\cdots$, hold with
remarkable accuracy.'\cite{Einstein23} Namely,   all gravitational
effects can be ignored on Einstein's `Galilean regions' in such a
way that his \SR\ with full Poincar\'e symmetry should hold {\it
locally}. This is
  because  all these regions are {\it
  finite}. Although in practice, it may still be regarded as {\it
global symmetry} approximately {\it with remarkable accuracy}.

If there are two such kind of `finite
  regions' of full local Poincar\'e invariance at different but
  nearby
 positions, how to pass from one  to another?

 According to Einstein, there should be gravity
 in-between
 these `regions'. Therefore, in order to transit from one to another, some curved
 spacetime with gravity in-between should be passed. In other words, in order to
 connect  these
  `regions' together,
 some gravitational field as
 interaction  should be taken into account. Since there is local Poincar\'e symmetry
 in  these `regions', in order to transit in-between, the spacetime with gravity should
 also be of  local Poincar\'e symmetry!
 Otherwise, it
 cannot be consistently transited from one `region' to another
 if Poincar\'e  symmetry cannot be maintained  locally
 in the course of transition. For any number of such `finite
 regions', it is the same.

This may also be seen from another angle in terminology  of
differential geometry. Each of {\it finite} `Galilean regions' is
essentially a portion of
 a \Mink-space with Poincar\'e symmetry isomorphic to an $R^4$, so
 that
 there are intersections among these
  \Mink-spaces with different  `finite regions' at different positions. And the transition functions on these
 intersections  should also be
 valued in Poincar\'e symmetry. Further,
 these \Mink-spaces with `finite regions' may be
 viewed as tangent spaces  at different positions of a curved
  manifold as the spacetime with
 gravity and the transition functions are valued in {\it local} Poincar\'e symmetry.

 Thus, it seems to be the core of  Einstein's idea on gravity
 that the theory of gravity should be based on the
localization of his \SR\ with Poincar\'e group as full symmetry
anywhere and anytime on some curved spacetimes. For the sake of
definiteness, we name this principle as the  localized \PoI\, with
full local symmetry or {\it the principle of localization}.
   Mathematically, this indicates
 that  spacetimes with gravity might be such a kind of manifolds that  on them the
  \Mink-space with (local) full Poincar\'e symmetry should be as a kind of tangent
 spaces anywhere and anytime in the universe.
 If so, the \PoI\, as a benchmark
 should be
 localized on the spacetimes with gravity and this should be in consistency with
 the case
 of the \Mink-space as a free spacetime
 where gravity might be ignored.

 But, in \GR, it is not really the
 case as was mentioned at beginning.

Due to the asymptotic behavior of the universe and in the light of
Einstein's `Galilean regions' as well as in view of  the  \dS\,
\SR, we may require that gravity in the universe should be based
on the localization of the  \dS\, \SR\, with localized \PoI\, in
local \dS-frame anywhere and anytime in the universe. Further, its
dynamics should also be properly of local \dS-invariance
characterized by a dimensionless constant $g\simeq (\Lambda G\hbar
/c^{3})^{1/2}\sim 10^{-61}$ from the cosmological constant
$\Lambda$ and the Planck length (see, e.g.\cite{lu80, duality}).
If so, the benchmark of physics is either the \PoI\, on the
\dS-space as a free space on the cosmic scale or its localization
with local \dS-invariance anywhere and anytime in the universe. In
addition, the evolution of the universe can also fix on the local
inertial frames of \dS-invariance in the same manner as the case
without gravity or where gravitational effects can be ignored at
very high accuracy.

A simple model for the \dS-gravity has implied that these points
should work.

In fact, from Cartan connection 1-form $\theta^{ab}=B^{ab}_{~~j}dx^j
\in \mathfrak{so}(1,3)$ and  Lorentz frame 1-form $\theta^a=e^a_j
dx^j$ on  Riemann-Cartan manifold of Einstein-Cartan theory
\cite{Cartan, EC, trautm}, it follows a kind of
connections valued at \dS-algebra \cite{dSG, hyg76, T77}%
\be\label{dSLB}%
{\cal{B}}:={\cal{B}}_jdx^j, ~ {\cal{B}}_j:=(
{\cal{B}}^{AB}_{~~j})_{A,B=0,\cdots, 4}= \left(
\begin{array}{cc}
B^{ab}_{~~j} & R^{-1} e^a_j\\
-R^{-1}e^b_{j} &0
\end{array}
\right ) \in \mathfrak{so}(1,4). %
\ee%
The curvature valued at \dS-algebra reads:
\be\label{dSLF}%
{\cal{F}}_{jk}= ( {\cal{F}}^{AB}_{~~jk})&=&\left(
\begin{array}{cc}
F^{ab}_{~~jk} + 2R^{-2}e^{ab}_{~~ jk} & R^{-1} T^a_{~jk}\\
-R^{-1}T^b_{~jk} &0
\end{array}
\right ) \in \mathfrak{so}(1,4),
\ee%
where $e^a_{~bjk}=\frac{1}{2}(e^a_je_{bk}-e^a_ke_{bj}),
e_{bj}=\eta_{ab}e^a_j$, $ F^{ab}_{~~ jk}$ and $ T^a_{~jk}$ are
curvature and torsion of Cartan connection.\omits{
\be\omits{\label{T2form}%
\Omega^a&=&d\theta^a+\theta^a_{~b} \wedge
\theta^b=\frac{1}{2}T^a_{~jk}dx^j\wedge dx^k\\\label{F2form}
\Omega^a_{~b}&=&d\theta^a_{~b}+\theta^a_{~c}\wedge\theta^c_{~b}=\frac{1}{2}F^a_{~b
jk}dx^j\wedge dx^k;\\}\label{Ta}
T^a_{~jk}&=&\partial_je^a_k-\partial_ke^a_j+B^a_{~c j}e^c_k-B^a_{~c
k}e^c_j,\\\label{Fab}%
F^a_{~b jk}&=&\partial_jB^a_{~bk} -\partial_kB^a_{~bj}+B^a_{~cj}B^c_{~bk}-B^a_{~ck}B^c_{~bj}.%
\ee%
Here $B^a_{~bj}=\eta_{ac}B^{ac}_{~~j}$ and so on. They satisfy the corresponding Bianchi identities.%
\be\label{BianchiL1}
d\Omega^a&=&\Omega^a_{~c}\wedge\theta^c-\theta^a_{~c}\wedge
\Omega^c,\\\label{BianchiL2} %
d\Omega^a_{~b}&=&d\Omega^a_{~c}\wedge
\theta^c_{~b}-\theta^a_{~c}\wedge\Omega^c_{~b}. \ee}%

The total action of the model
with source may be taken as%
\be\label{S_t}%
S_T=S_{GYM}+S_m,%
\ee%
where $S_m$ is the  action of  source  and $S_{GYM}$ the
Yang-Mills-like action of gravity: %
\be\nno%
S_{GYM}&=&\frac{1}{4g^2}\int_{M}d^4x { e}
{\bf Tr}_{dS}({\cal F}_{jk}{\cal F}^{jk})\\
&=& \int_{M}d^4x {
e}\left[\chi(F+2\Lambda)-\frac{1}{4g^2}F^{ab}_{~jk}F_{ab}^{~jk} {
+}\frac{\chi}{2} T^a_{~jk}T_a^{~jk}\right].\label{GYM}
\ee%
Here $e=\det(e^a_j)$, a dimensionless constant $g$ should be
introduced as usual in  gauge theory to describe the
self-interaction of the gravitational field, $\chi$ a dimensional
coupling constant related to $g$ and $R$, and $F={
\frac{1}{2}}F^{ab}_{~jk}e_{ab}^{~jk}$ the scalar curvature of Cartan
connection, the same as the action in  Einstein-Cartan theory. In
order to make sense in comparison  with  Einstein-Cartan theory, we
take $\chi=1/(8\pi G)$ and $g^{-2}\simeq 3\chi\Lambda^{-1}$ with
$\hbar=c=1$. In fact, $g^2\simeq G\hbar c^{-3}\Lambda$.\omits{, the
same as the one introduced in the last section in the sense of the
Planck scale-$\Lambda$ duality. This is why we have used the same
symbol in the different cases.}

Although the gravitational field equation now should be of
Yang-Mills type, this model does pass the observation tests in
solar-scale and there are
 simple cosmic models  having `Big
Bang'. But, different from \GR, there are `energy-momentum-like
tensors' for gravity from the $F^2$ and $T^2$ terms  as a kind of
the `dark stuffs' in the action (\ref{GYM}). In fact, by means of
the relation between Cartan connection $B^{ab}_{~j}$ and  Ricci
rotational coefficients $\gamma^{ab}_{~j}$, we may pick up
Einstein's action from Einstein-Cartan's action $F$, and the rest
terms in (\ref{GYM}) are all `dark stuffs' in view of \GR. Thus,
this model should provide an alternative framework for the cosmic
data analysis.

In this model, there is the cosmological constant $\La$ from local
\dS-symmetry so that it is not just a `dummy' constant at classical
level as in \GR. In fact, this model can be viewed as a kind of
\dS-gravity in a `special gauge' and the 4-dimensional
Riemann-Cartan manifolds should be a kind of 4-dimensional umbilical
manifolds that there is local \dS-spacetime together with `gauged'
\dS-algebra anywhere and anytime (see, e.g. \cite{hyg76, lu80,
duality}).

It is interesting that the model is renormalizable \cite{QG} with an
$SO(5)$ gauge-like Euclidean action having a Riemann sphere as an
instanton. Thus,  quantum tunneling scenario may support
$\Lambda>0$. For the gauge-like gravity, asymptotic freedom  may
indicate that the coupling constant $g$ \omits{should be running to
zero at infinity momentum. For gravity, however,  , so $g$ }should
be very tiny and link  the cosmological constant $\Lambda$ with the
Planck length $\ell_P$ properly, since both the $\Lambda$ and Planck
scale as
  fixed points provide an infrared and an ultraviolet cut-off, respectively.% for the momentum.

We will explain these issues in detail elsewhere.
%%%%%%%%%%%%%%%%%%%%%%%%%%%%%%%%%%%%%%%%%%%%%%%%%%%%%%%%%%%%%%%%%%%%%%%%
%%%%%%%%%%%%%%%%%%%%%%%%       Section 6       %%%%%%%%%%%%%%%%%%%%%%%%
%%%%%%%%%%%%%%%%%%%%%%%%%%%%%%%%%%%%%%%%%%%%%%%%%%%%%%%%%%%%%%%%%%%%%%%%
%%%%%%%%%%%%%%%%%%%%%%%%%%%   conclusion %%%%%%%%%%%%%%%%%%%%%%%%%%%%%
\section{Concluding remarks}

 In physics of the last century, symmetry,  localization of  symmetry and symmetry breaking play
very important roles. For the cosmic scale physics without or with
gravity, it should be also the case. In  view of the  \dS\, \SR\,
and in the light of Einstein's `Galilean regions', the \PoI\, with
maximum symmetry and its localization should still play a central
role as the benchmarks of physics in the large scale.

If the universe is asymptotic to a \dS-space, it should be
asymptotic to a slightly closed Robertson-Walker-like \dS-space,
which closely relates to the \BdS-space with the \PoI. Therefore,
the evolution of the universe  also supports the \PoI\, on the
\BdS-space and fix on the Beltrami systems without Einstein's
`argument in a circle'. Thus, the \PoI\, of the \dS\, \SR\ is a
benchmark of physics on the cosmic scale when gravity can be
ignored.

We may require that on the spacetimes with gravity there should be
locally  the \PoI\, with local inertial frames of full
\dS-symmetry anywhere and anytime. Then,  the evolution of the
universe can also fix on these local inertial frames.  A simple
model for the \dS-gravity has implied these requirements.

 Thus, the \PoI\, of the  \dS\, \SR\, and its
localization are  consistent benchmarks of physics without or with
gravity in the universe.

%%%%%%%%%%%%%%%%%%%%%%%%%%%%%%%%%%%%%%%%%%
%%%%%%%%% acknowledgments %%%%%%%%%%%%%%%%
%%%%%%%%%%%%%%%%%%%%%%%%%%%%%%%%%%%%%%%%%

\begin{acknowledgments}
 We are grateful  to   Z. Chang, C.-G. Huang, W.L. Huang, Q.K. Lu, J.Z. Pan, Y. Tian, Z.
Xu, X. Zhang, B. Zhou, C.J. Zhu and Z.L. Zou for valuable
discussions and comments. This work is partly supported by NSFC
under Grant 90503002.
\end{acknowledgments}

\end{document}